\documentclass[12pt]{article}
\pdfoutput=1
\usepackage{jheppub}

\usepackage{amsmath,amssymb,amsfonts}
\usepackage{graphicx}
\usepackage{color}
\usepackage{tikz}
\usetikzlibrary{calc}
\usepackage{dsfont}

\usepackage{subfigure}

\def\cA{{\cal A}}
\def\cB{{\cal B}}
\def\cC{{\cal C}}

\def\cG{{\cal G}}

\def\cY{{\cal Y}}

\def\cS{{\cal S}}

\def\cM{{\cal M}}
\def\q{{\mathfrak q}}

\DeclareMathOperator{\vol}{vol}

\makeatletter\def\l@subsubsection#1#2{}%
\makeatother

\usepackage{textcomp}

\def\Im{\mathop{\rm Im}}
\def\Re{\mathop{\rm Re}}
\def\RR{\mathds{R}}
\def\CC{\mathds{C}}
\def\ZZ{\mathds{Z}}

\begin{document}

\title{On minimal Type IIB $AdS_6$ solutions with commuting 7-branes}

\author{Andrea Chaney and}
\emailAdd{achaney@physics.ucla.edu}
\author{Christoph F.~Uhlemann} 
\emailAdd{uhlemann@physics.ucla.edu}

\affiliation{Mani L.\ Bhaumik Institute for Theoretical Physics\\
Department of Physics and Astronomy\\
University of California, Los Angeles, CA 90095, USA}

\abstract{
We construct Type IIB supergravity solutions with geometry $AdS_6\times S^2$ warped over a disc with two boundary points where 5-branes emerge and punctures with 7-brane monodromy. They describe $(p,q)$ 5-brane junctions with two groups of like-charged external 5-branes that are unconstrained by the {\it s}-rule and an additional group of constrained 5-branes. The dual 5d SCFTs include various theories discussed previously in the literature. We match SCFT operators with scaling dimension of $\mathcal O(N)$ with their representation in supergravity to support the proposed dualities.
}

\maketitle

\section{Introduction and summary}

Interacting five-dimensional superconformal field theories (SCFTs) do not have a conventional Lagrangian description, but they can be engineered in string and M-theory, and they often admit relevant deformations that flow to Lagrangian gauge theories in the infrared \cite{Seiberg:1996bd,Morrison:1996xf,Douglas:1996xp,Intriligator:1997pq,Jefferson:2017ahm,Jefferson:2018irk,Xie:2017pfl}. A particularly versatile approach to constructing 5d SCFTs is via $(p,q)$ 5-brane webs in Type IIB string theory \cite{Aharony:1997ju,Kol:1997fv,Aharony:1997bh}, and the constructions can be further generalized by including 7-branes \cite{DeWolfe:1999hj}. General brane webs engineer mass deformations of the 5d SCFTs, while the conformal limits are described by junctions of 5-branes at a point.

In the absence of a Lagrangian description for the 5d SCFTs, AdS/CFT dualities can be particularly useful for gaining insight into these theories. A large class of Type IIB supergravity solutions describing the near-horizon limit of $(p,q)$ 5-brane junctions has been constructed in   \cite{D'Hoker:2016rdq,DHoker:2016ysh,DHoker:2017mds},\footnote{Previous analyses of the BPS equations can be found in \cite{Apruzzi:2014qva,Kim:2015hya,Kim:2016rhs}. T-duals of a Type IIA solution \cite{Brandhuber:1999np}, which has been studied extensively \cite{Bergman:2012kr,Jafferis:2012iv,Bergman:2012qh,Assel:2012nf,Bergman:2013koa,Pini:2014bea,Chang:2017mxc,Passias:2018swc}, have been discussed in \cite{Lozano:2012au,Lozano:2013oma}.} and provides the stepping stone for AdS/CFT analyses of the 5d SCFTs engineered in Type IIB string theory. The geometry takes the form $AdS_6\times S^2$ warped over a disc $\Sigma$. The boundary of $\Sigma$, at which the $S^2$ collapses to smoothly close off the ten-dimensional geometry, contains isolated points at which the 5-branes of the associated 5-brane junction emerge. Various aspects of these solutions have been studied since \cite{Gutperle:2017tjo,Kaidi:2017bmd,Gutperle:2018wuk,Hong:2018amk,Malek:2018zcz}, and detailed comparisons to field theory results support the identification of these solutions as holographic duals for 5d SCFTs engineered by $(p,q)$ 5-brane junctions \cite{Bergman:2018hin,Fluder:2018chf}. 

The solutions have been extended to include punctures on $\Sigma$ with commuting 7-brane monodromies in \cite{DHoker:2017zwj}.\footnote{More precisely, if multiple punctures are present the $SL(2,\RR)$ matrices describing the monodromies are required to commute.} 
These solutions with monodromy correspond to 5-brane junctions where, within one group of like-charged 5-branes, multiple 5-branes terminate on the same 7-brane \cite{Gutperle:2018vdd,Bergman:2018hin}, and are thus constrained by the $s$-rule \cite{Benini:2009gi}. The 5d SCFTs engineered by such constrained junctions are related to those engineered by unconstrained junctions through RG flows along the Higgs branch. When considering mass deformations of the SCFT, such that the 5-brane junction becomes a 5-brane web with open faces, the 7-branes on which multiple 5-branes end may each be moved into the web. Attached 5-branes are successively removed through the (inverse) Hanany-Witten effect in the process, and the 7-branes may be moved to a face where they have no 5-branes attached. The conformal limit of this form of the brane web is described by the supergravity solutions with monodromy. The locations of the 7-branes inside the brane web are encoded in the positions of the punctures on $\Sigma$.\footnote{The T-duals of the Type IIA solution can be understood to arise from solutions with 7-branes upon smearing the 5- and 7-branes \cite{Lozano:2018pcp}.}

Solutions without monodromy have at least three points on the boundary of $\Sigma$ where groups of like-charged external $(p,q)$ 5-branes emerge, corresponding to the fact that three groups of 5-branes are needed to create an intersection point and thus a 5d SCFT. The aforementioned understanding of solutions with monodromy suggests that solutions with punctures should exist with only two boundary points where groups of unconstrained external 5-branes emerge, while an additional group of constrained 5-branes is realized by punctures. In this note we construct such solutions, following the strategy of \cite{DHoker:2017zwj} but relaxing one of the regularity assumptions. These solutions are minimal in the sense that the number of groups of external 5-branes can not be smaller for this class of solutions with only commuting monodromies.

The dual field theories include a variety of theories that have been discussed previously in the literature. In particular, a class of 5d $T_N$ theories \cite{Benini:2009gi}, which reduce to 4d $T_N$ theories upon compactification. The types of punctures characterizing the 4d $T_N$ theories are encoded in the 5d version in the combinatorics of how the external 5-branes terminate on 7-branes. Holographic duals for the 5d $T_N$ theories corresponding to three maximal partitions have been discussed in \cite{Bergman:2018hin}. With the solutions constructed here this can be generalized to 5d $T_N$ theories with two maximal and one arbitrary partition. This includes the $R_{0,N}$ theories of \cite{Chacaltana:2010ks,Bergman:2014kza} and the $\chi_N^k$ theories of \cite{Bergman:2014kza}, and in particular large-$N$ generalizations of the $E_7$ theory.

For two subclasses of two-pole solutions we discuss in detail the associated 5-brane junctions and dual 5d SCFTs. The first is a subclass of the $T_N$ theories, which includes the $R_{0,N}$ and $\chi_N^k$ theories and the $E_7$ theory, and the second one is a constrained version of the D5/NS5 intersection discussed initially in \cite{Aharony:1997bh}, which includes the $S_5$ theory of \cite{Bergman:2018hin} as well as the $E_6$ and $E_7$ theories. We discuss relevant deformations of the SCFTs that flow to quiver gauge theories, in which we identify ``stringy'' operators with scaling dimensions of $\mathcal O(N)$ in the large-$N$ limit and determine their scaling dimensions. This data can be compared to results obtained from the proposed dual supergravity solutions, where the stringy operators are realized as strings. We show that the two approaches agree, to support the proposed dualities.

The paper is organized as follows. In sec.~\ref{sec:review} we review the Type IIB $AdS_6$ solutions. In sec.~\ref{sec:min-sol} we construct solutions with two boundary points where 5-branes emerge and an arbitrary number of punctures with commuting monodromies. In sec.~\ref{sec:5d-SCFTs} we discuss the dual field theories for two subclasses of solutions and match supergravity results to field theory.

\section{Warped \texorpdfstring{$AdS_6$}{AdS6} solutions with 7-branes}\label{sec:review}
The geometry in the solutions of \cite{D'Hoker:2016rdq,DHoker:2016ysh,DHoker:2017mds,DHoker:2017zwj} is a warped product of $AdS_6\times S^2$ over a Riemann surface $\Sigma$, with metric and complex two-form given by
\begin{align}\label{eqn:ansatz}
 ds^2 &= f_6^2 \, ds^2 _{\mathrm{AdS}_6} + f_2 ^2 \, ds^2 _{\mathrm{S}^2} + 4\rho^2 dw d\bar w~,
 &
 C_{(2)}&=\cC \vol_{S^2}~,
\end{align}
where $w$ is a complex coordinate on $\Sigma$ and $\vol_{S^2}$ the canonical volume form on a unit-radius $S^2$.
The general solution to the BPS equations is parametrized by two locally holomorphic functions $\cA_\pm$ on $\Sigma$. 
The metric functions are
\begin{align}\label{eqn:metric}
f_6^2&=\sqrt{6\cG T}~,
&
f_2^2&=\frac{1}{9}\sqrt{6\cG}\,T ^{-\tfrac{3}{2}}~,
&
\rho^2&=\frac{\kappa^2}{\sqrt{6\cG}} T^{\tfrac{1}{2}}~,
\end{align}
where
\begin{align}\label{eq:kappa-G}
 \kappa^2&=-|\partial_w \cA_+|^2+|\partial_w \cA_-|^2~,
 &
 \partial_w\cB&=\cA_+\partial_w \cA_- - \cA_-\partial_w\cA_+~,
 \\
 \cG&=|\cA_+|^2-|\cA_-|^2+\cB+\bar{\cB}~,
 &
  T^2&=\left(\frac{1+R}{1-R}\right)^2=1+\frac{2|\partial_w\cG|^2}{3\kappa^2 \, \cG }~.
  \label{eqn:Gdef}
\end{align}
The function $\cC$ and the axion-dilaton scalar $B=(1+i\tau)/(1-i\tau)$ are given by
\begin{align}\label{eqn:flux}
 \mathcal{C} & =  \frac{2i}{3}\left(
 \frac{\partial_{\bar w}\cG\partial_w\cA_++\partial_w \cG \partial_{\bar w}\bar\cA_-}{3\kappa^{2}T^2} - \bar{\mathcal{A}}_{-} - \mathcal{A}_{+}  \right)
~,
\\
B &=\frac{\partial_w \cA_+ \,  \partial_{\bar w} \cG - R \, \partial_{\bar w} \bar \cA_-   \partial_w \cG}{
R \, \partial_{\bar w}  \bar \cA_+ \partial_w \cG - \partial_w \cA_- \partial_{\bar w}  \cG}~.
\end{align}

Physically regular solutions without monodromy were constructed in \cite{DHoker:2016ysh,DHoker:2017mds} for the case where $\Sigma$ is a disc or equivalently the upper half plane. On the upper half plane the locally holomorphic functions are
\begin{align}\label{eqn:cA-0}
 \cA^s_\pm  &=\cA_\pm^0+\sum_{\ell=1}^L Z_\pm^\ell \ln(w-r_\ell)~,
\end{align}
with a superscript $s$ indicating that they are single-valued in the interior of $\Sigma$ and $\cA_+^0=-\overline{\cA_-^0}$.
The differentials $\partial_w\cA_\pm^s$ have poles at isolated points $r_\ell$ on the real line, and non-degenerate solutions need $L\geq 3$.
The residues  $Z_\pm^\ell$ are given in terms of $L-2$ complex parameters $s_n$ in $\Sigma$ and a complex normalization $\sigma$ by
\begin{align}\label{eqn:residues}
Z_+^\ell  &=
 \sigma\prod_{n=1}^{L-2}(r_\ell-s_n)\prod_{k \neq\ell}^L\frac{1}{r_\ell-r_k}~,
 &
 Z_-^\ell&= - \overline{Z_+^\ell}~.
\end{align}
There are additional regularity conditions whose precise form will not be needed here.

Physically regular solutions with monodromy were constructed in \cite{DHoker:2017zwj}. We will, without loss of generality, restrict the discussion to the case of D7-brane monodromy. The additional parameters are the locations of the punctures, $w_i$, the orientation of the associated branch cuts, parametrized by complex phases $\gamma_i$ with $|\gamma_i|=1$, and relative weights $n_i$. With
\begin{align}\label{eq:f-def}
f(w) &= \sum _{i=1}^I \frac{n_i^2}{4\pi} \ln \left ( \gamma_i\,\frac{ w-w_i}{w -\bar w_i} \right )~,
\end{align}
the locally holomorphic functions for a solution with D7-brane monodromy are 
\begin{align}\label{eqn:cA-monodromy}
 \cA_\pm&= \cA_\pm^s + \int_\infty^w dz \;f(z)\sum_{\ell=1}^L \frac{Y^\ell}{z-r_\ell}~,
 &
 Y^\ell&=Z_+^\ell-Z_-^\ell~,
\end{align}
where the integration contour is chosen such that it does not cross any branch cuts.
The residues of $\partial_w\cA_\pm$ at the poles $r_\ell$ are 
\begin{align}\label{eq:cY}
\cY_\pm^\ell&\equiv Z_\pm^\ell+ f(r_\ell)Y^\ell~.
\end{align}
The remaining regularity conditions constraining the parameters are
\begin{subequations}\label{eq:reg-gen}
\begin{align}
\label{eq:w1-summary}
 0&=2\cA_+^0-2\cA_-^0+\sum_{\ell=1}^LY^\ell \ln|w_i-r_\ell|^2~,
 &i&=1,\cdots,I~,
 \\
 0&=2\cA_+^0\cY_-^k-2\cA_-^0\cY_+^k
 +\sum_{\ell\neq k} Z^{[\ell, k]}\ln |r_\ell-r_k|^2+Y^kJ_k~,
 &
 k&=1,\cdots,L~,
 \label{eq:DeltaG0-summary}
\end{align}
\end{subequations}
where $Z^{[\ell, k]}\equiv Z_+^\ell Z_-^k-Z_+^k Z_-^\ell$.
With $\cS_k\subset\lbrace 1,\cdots, I\rbrace$ the set of branch points for which the associated branch cut intersects the real line in the interval $(r_k,\infty)$, $J_k$ is given by
\begin{align}\label{eq:Jk-def}
 J_k&=\sum_{\ell=1}^L Y^\ell\Bigg[\int_\infty^{r_k} dx f^\prime(x)  \ln |x-r_\ell|^2
 +\sum_{i\in\cS_k} \frac{i n_i^2}{2} \ln |w_i-r_\ell|^2\Bigg]~,
\end{align}
where the integral over $x$ is along the real line.
At the poles $r_\ell$, $(p_\ell,q_\ell)$ 5-branes emerge, with charges given in terms of $\cY_+^\ell$ by \cite{Bergman:2018hin}
\begin{align}\label{eq:cY-pq}
 \cY_+^\ell&=\frac{3}{4}\alpha^\prime \left(q_\ell+i p_\ell\right)~,
\end{align}
where a D5 brane has charge $(\pm 1,0)$ and an NS5 brane $(0,\pm 1)$.

\section{Minimal solutions with commuting 7-branes}\label{sec:min-sol}

In this section we construct 2-pole solutions with punctures and $SL(2,\RR)$ monodromy, and explicitly solve all regularity conditions.

\subsection{Regularity conditions}

The construction of solutions with monodromy, as summarized in sec.~\ref{sec:review}, starts out from solutions without monodromy. The conditions in  (\ref{eq:reg-gen}) are sufficient to guarantee the regularity conditions  $\kappa^2, \cG > 0$
in the interior of $\Sigma$ and $\kappa^2=\cG=0$ on $\partial\Sigma$.
This was shown in \cite{DHoker:2017zwj} using that $\cA_\pm^s$ in (\ref{eqn:cA-monodromy}) correspond to regular solutions without monodromy. 
In particular, $\kappa^2$ for solutions with monodromy can be expressed as
\begin{align}\label{eq:kappa2}
 \kappa^2
 &=\kappa_s^2 -(f+\bar f)\left|\partial_w\cA_+^{s}-\partial_w\cA_-^{s}\right|^2~,
 &
 \kappa_s^2&=-|\partial_w\cA_+^{s}|^2+|\partial_w\cA_-^{s}|^2~.
\end{align}
If $\kappa^2_s$ is positive in the interior of $\Sigma$ and zero on the boundary, which requires $L\geq 3$ poles in $\partial_w\cA_\pm^s$, the properties of $f$ guarantee that the same is true for $\kappa^2$, thus satsifying the regularity conditions.

However, while positivity of $\kappa_s^2$ in the interior of $\Sigma$ and its vanishing on the boundary are sufficient to guarantee regularity of $\kappa^2$, they are not both necessary: One may allow for $\kappa_s^2$ to vanish identically. The properties of $f$ still guarantee that $\kappa^2=0$ on $\partial\Sigma$. If $\partial_w\cA_+^{s}-\partial_w\cA_-^{s}$ is non-zero throughout $\Sigma$, $\kappa^2$ is also positive in the interior of $\Sigma$. This permits the construction of 2-pole solutions. The discussion of the remaining regularity conditions proceeds analogously to the case with three or more poles, and leads to the conditions (\ref{eq:reg-gen}) with $L=2$.

\subsection{\texorpdfstring{$\rm SL(2,\RR)$}{SL(2,R)} automorphisms of \texorpdfstring{$\mathbb{H}$}{H}}\label{sec:sl2r}
The explicit expressions in the constructions to follow can be simplified by a convenient choice of coordinate on $\Sigma$.
The $\rm SL(2,\RR)$ automorphisms of the upper half plane act as
\begin{align}
 w&\mapsto \frac{aw+b}{cw+d}~,
 & ad-bc&=1~.
\end{align}
They can be used to fix the position of both poles, and we choose them symmetric with respect to reflection across the imaginary axis as
\begin{align}
\label{eq:poles}
 r_1&=1~,& r_2&=-1~.
\end{align}
This leaves a family of residual $\rm SL(2,\RR)$ transformations.
Namely, those with $c=b$, $d=a$ and $a^2=1+b^2$.
They can be used to map an arbitrary point in the interior of the upper half plane to the imaginary axis.

\subsection{Ansatz}

Since the residues as given in (\ref{eqn:residues}) sum to zero by construction, for $L=2$ we have
\begin{align}\label{eq:Zp12}
 Z_\pm\equiv Z_\pm^1=-Z_\pm^2~, &
 &
 Z_+&=-\bar Z_-~.
\end{align}
The differentials $\partial_w\cA^s_\pm$  are given by
\begin{align}
 \partial_w\cA_\pm^s&=\frac{2Z_\pm}{w^2-1}~.
\end{align}
This leads to identically vanishing $\kappa_s^2$ and a solution without monodromy would be degenerate. For solutions with non-trivial monodromy, however, $\kappa^2$ as given in (\ref{eq:kappa2}) is positive in the interior of $\Sigma$, provided that
\begin{align}\label{eq:no-D5}
 Z_+\neq Z_-~.
\end{align}
With the poles as in (\ref{eq:poles}), the locally holomorphic functions (\ref{eqn:cA-monodromy}) take the form
\begin{align}
 \cA_\pm&=\cA_\pm^0+Z_\pm\ln\frac{w-1}{w+1}+2(Z_+- Z_-)\int_\infty^w dz\,\frac{f(z)}{z^2-1}~,
\end{align}
with $f$ given in (\ref{eq:f-def}). 
We use the residual $\rm SL(2,\RR)$ transformations discussed in sec.~\ref{sec:sl2r} to fix, without loss of generality, one puncture to lie on the imaginary axis,
\begin{align}\label{eq:fix-w1}
 w_1&=i\alpha_1~,& \alpha_1\in\RR^+~.
\end{align}

\subsection{Solving the regularity conditions}

The remaining regularity conditions are given in (\ref{eq:reg-gen}).
The first set of conditions, (\ref{eq:w1-summary}), for $L=2$ and the choice of poles in (\ref{eq:poles}) becomes
\begin{align}\label{eq:w1}
  0&=2\cA_+^0-2\cA_-^0+(Z_+-Z_-) \ln\left|\frac{w_i-1}{w_i+1}\right|^2~,
 &i&=1,\ldots,I~.
\end{align}
With $w_1$ fixed as in (\ref{eq:fix-w1}), the $I=1$ condition implies $\cA_+^0=\cA_-^0$. 
In view of $\cA_+^0=-\overline{\cA_-^0}$, the constant $\cA^0_+$ therefore has to be purely imaginary.
The remaining conditions in (\ref{eq:w1}) reduce to
\begin{align}
0&=\ln\left|\frac{w_i-1}{w_i+1}\right|^2~,
 &i&=2,\ldots,I~. 
\end{align}
This implies that all punctures have to be on the imaginary axis. They may be parametrized more conveniently as
\begin{align}\label{eq:branch-points}
w_i&=i\alpha_i~,&\alpha_i&\in\RR^+~.
\end{align}
It remains to solve the second set of conditions, (\ref{eq:DeltaG0-summary}). Due to (\ref{eq:Zp12}),  $Z^{[1,2]}=0$.
Using that $\cA_+^0=\cA_-^0$, the definition of $\cY_\pm^\ell$, and that $Y^k\neq 0$ due to (\ref{eq:no-D5}), the conditions reduce to
\begin{align}\label{eq:G0-sol}
 \cA_+^0&=\frac{1}{2}J_1~,
 &
 J_1&=J_2~.
\end{align}
Since the $J_k$ are imaginary by construction, the first condition simply fixes $\cA_+^0$.
The second term in the definition of $J_k$ in (\ref{eq:Jk-def}) vanishes for $L=2$.
Explicit evaluation, using (\ref{eq:branch-points}), yields
\begin{align}
 J_k&=(Z_+-Z_-)\int_\infty^{r_k} dx f^\prime(x)\ln\left|\frac{x-1}{x+1}\right|^2~.
\end{align}
The integrand is odd under $x\rightarrow -x$, which is sufficient to show that $J_1-J_2=0$, such that the second condition in (\ref{eq:G0-sol}) is satisfied.

\subsection{Summary and parameter count}

In summary, for an arbitrary choice of $\alpha_i\in\RR^+$, $n_i\in\RR$ and $\gamma_i$ with $|\gamma_i|=1$, for $i=1,..,I$, as well as a complex $Z_+$ with non-vanishing real part, there is a regular 2-pole solution.
The locally holomorphic functions are
\begin{align}
 \cA_\pm&=\cA_\pm^0+Z_\pm\ln\frac{w-1}{w+1}+2Y\int_\infty^w dz\,\frac{f(z)}{z^2-1}~,
 &
 Y&\equiv Z_+- Z_-~,
\end{align}
with $Z_-=-\bar Z_+$ and
\begin{align}
 f(w) &= \sum _{i=1}^I \frac{n_i^2}{4\pi} \ln \left ( \gamma_i\,\frac{ w-i\alpha_i}{w +i\alpha_i} \right )~,
 &
 \cA_\pm^0&=Y\int_\infty^{1} dx f^\prime(x)\ln\left|\frac{x-1}{x+1}\right|\,.
\end{align}
The residues at the poles, which translate to the charges of the external 5-branes, are
\begin{align}\label{eq:cY-2pole}
 \cY_+^1&=Z_+ + Yf(1)~,
 &
 \cY_+^2&=-Z_+ - Yf(-1)~.
\end{align}
The requirement for $Z_+$ to have non-vanishing real part corresponds to the fact that a brane configuration with only D5 and D7 branes does not realize a 5d theory.

For the construction we have assumed punctures with D7-brane monodromy.
That is, the $\rm SL(2,\RR)$ monodromy around the punctures at $w_i=i\alpha_i$ is given by
\begin{align}
 M_{[n_i,0]}&=\begin{pmatrix} 1 & n_i^2\\0 & 1\end{pmatrix}~.
\end{align}
A solution with generic but commuting $\rm SL(2,\RR)$ monodromies can be obtained by an $\rm SL(2,\RR)\cong SU(1,1)$ transformation parametrized by $u,v\in\CC$ with $|u|^2-|v|^2=1$ \cite{DHoker:2017zwj}
\begin{align}
\cA_+ & \to   u \cA_+ - v \cA_- ~,
&
\cA_- & \to     - \bar v \cA_+ + \bar u \cA_- ~.
\end{align}
An appropriate choice of transformation is
\begin{align}\label{eq:uqvq}
 u&=\frac{1+\eta_+\eta_-}{2\eta_-}~,
 &
 v&=\frac{1-\eta_+\eta_-}{2\eta_-}~,
 &
 \eta_\pm&=p\mp iq~.
\end{align}
This introduces another independent parameter for the general solution, which is the ratio of $p$ and $q$. 
The $\rm SL(2,\RR)$ automorphisms of the upper half plane are entirely fixed by the choices of poles in (\ref{eq:poles}) and of the first puncture in (\ref{eq:fix-w1}).
We thus recover a total of $3+3I$ parameters for a solution with $I$ punctures.

\section{5-brane junctions and 5d SCFTs}\label{sec:5d-SCFTs}

We discuss two classes of 5d SCFTs. The holographic duals for the first class are generically three-pole solutions which reduce to two-pole solutions for particular choices of the parameters. The holographic duals for the second class are generically two-pole solutions. We discuss field theory operators which can be identified in the holographic duals and show that their properties match in the two descriptions.

\begin{figure}
\centering
\subfigure[][]{\label{fig:TN-web-2}
\begin{tikzpicture}[scale=0.95]
    \foreach \i in {-3/2,-1/2,1/2,3/2}{
      \draw (-0.4,0.127*\i) -- (-1.8,0.127*\i);
      \draw (0.127*\i,-0.4) -- (0.127*\i,-1.8) [fill=black] circle (1pt);
      \draw (0.28+0.09*\i,0.28-0.09*\i) -- (1.27+0.09*\i,1.27-0.09*\i) [fill=black] circle (1pt) ;
    }
    \draw[fill=black] (-1.8,0.127) ellipse (1pt and 2.8pt);
    \draw[fill=black] (-1.8,-0.127) ellipse (1pt and 2.8pt);
    \draw[fill=gray] (0,0) circle (0.38);
    
    \node [anchor=east] at (-1.9,0) {\footnotesize $Y_1=[2,2]$};
    \node [anchor=south west] at (1.3,1.3) {\footnotesize $Y_2=[1^4]$};
    \node [anchor=north] at (0,-1.9) {\footnotesize $Y_3=[1^4]$};
\end{tikzpicture}
}\hskip 10mm
\subfigure[][]{\label{fig:TN-web-3}
\begin{tikzpicture}[scale=0.95]
    \foreach \i in {-2,-1,0,1,2}{
      \draw (-0.4,0.127*\i) -- (-1.8,0.127*\i);
      \draw (0.127*\i,-0.4) -- (0.127*\i,-1.8) [fill=black] circle (1pt);
      \draw (0.28+0.09*\i,0.28-0.09*\i) -- (1.27+0.09*\i,1.27-0.09*\i) [fill=black] circle (1pt) ;
    }
    \draw (-1.8,0.127*2) [fill=black] circle (1pt);
    \draw (-1.8,0.127*1) [fill=black] circle (1pt);
    \draw[fill=gray] (0,0) circle (0.38);
    \draw[fill=black] (-1.8,-0.127*1) ellipse (1pt and 4.4pt);

    \node [anchor=east] at (-1.9,0) {\footnotesize $Y_1=[3,1^2]$};
    \node [anchor=south west] at (1.3,1.3) {\footnotesize $Y_2=[1^5]$};
    \node [anchor=north] at (0,-1.9) {\footnotesize $Y_3=[1^5]$};
\end{tikzpicture}
}
\caption{On the left the $T_{N,K,j}$ junction with $N=4$, $K=j=2$, which realizes the $E_7$ theory. On the right hand side for $N=5$, $j=1$, $K=3$. \label{fig:TN-webs}}
\end{figure}

\subsection{The \texorpdfstring{$T_{N,K,j}$}{T-N-k-j} theory}

The 5d $T_N$ theories are realized by triple junctions of $N$ D5, $N$ NS5 and $N$ $(1,1)$ 5-branes \cite{Benini:2009gi}. They are characterized by the way in which the external 5-branes of the junction are partitioned into groups terminating on the same 7-brane. We discuss the case of unconstrained NS5 and $(1,1)$ 5-branes, but with $j$ groups of $K>1$ D5 branes each terminating on one D7-brane, leaving $N-jK$ unconstrained D5 branes (fig.~\ref{fig:TN-webs}). In the notation of \cite{Hayashi:2014hfa,Tachikawa:2015bga} for the partitions, this corresponds to
\begin{align}
 Y_1&=[K^j,1^{N-K j}]~,
 & Y_2&=Y_3=[1^N]~.
\end{align}
For $N=4$ and $j=K=2$ this is the $E_7$ theory \cite{Benini:2009gi}.
For $j=0$ or $K=1$ the $T_{N,K,j}$ theories reduce to the unconstrained $T_N$ theories, for which the supergravity duals were discussed in \cite{Bergman:2018hin}.
The $R_{0,N}$ theories of \cite{Chacaltana:2010ks,Bergman:2014kza} correspond to $K=N-2$, $j=1$, and the $\chi_N^k$ theories of \cite{Bergman:2014kza} are contained as $\chi_N^k=T_{N,N-k-1,1}$.
The $T_{N,K,j}$ theories are obtained from the unconstrained $T_N$ theories by RG flows along the Higgs branch, and the global symmetry is reduced from $SU(N)^3$ to
\begin{align}\label{eq:TN-symmetry}
 SU(N-j K)\times SU(j)\times SU(N)^2\times U(1)~. 
\end{align}

\begin{figure}
 \centering
\begin{tikzpicture}[scale=0.95]
    \foreach \i in {-2,-1,0,1,2}{
      \draw (-0.4,0.127*\i) -- (-2.6,0.127*\i);
      \draw (0.127*\i,-0.4) -- (0.127*\i,-1.8) [fill=black] circle (1pt);
      \draw (0.28+0.09*\i,0.28-0.09*\i) -- (1.27+0.09*\i,1.27-0.09*\i) [fill=black] circle (1pt);
    }
    \draw[fill=gray] (0,0) circle (0.38);

    \node [anchor=south west] at (1.3,1.3) {\footnotesize $N-K\,(-1,-1)$};
    \node [anchor=north] at (0,-1.9) {\footnotesize $N-K\,(0,1)$};
    
    \foreach \i in {-1,0,1}{   \draw (-3.4,0.127*\i) -- (-4.8,0.127*\i) [fill=black] circle (1pt); }
    \foreach \i in {-1/2,1/2}{
      \draw (-3+0.127*\i,-0.4) -- (-3+0.127*\i,-1.8) [fill=black] circle (1pt);
      \draw (-3-0.28-0.09*\i,0.28-0.09*\i) -- (-3-1.27-0.09*\i,+1.27-0.09*\i) [fill=black] circle (1pt);
    }
    \draw[fill=gray] (-3,0) circle (0.38);
    \node [anchor=north] at (-3,-1.9) {\footnotesize $K\,(0,1)$};
    \node [anchor=east] at (-4.9,0) {\footnotesize $N-jK\, (0,1)$};
    \node [anchor=south east] at (-4.3,1.3) {\footnotesize $K\,  (j-1,-1)$};
    
    \draw[fill] (-1.5,0.7) circle (1.5pt);
    \draw[dashed,thick] (-1.5,0.7) --(-1.5,2) node [anchor=south] {\small $T^j$};
    
    \draw[blue] plot [smooth] coordinates { (-4.8,0.127) (-3.6,0.9) (-2.55,1.0)  (-1.5,0.7)};
    
\end{tikzpicture}
\caption{Brane web obtained from the $T_{N,K,j}$ junction by mass deformation and 7-brane moves, with the  fundamental string expected to realize the meson operator shown in blue. The 5-brane numbers and charges are indicated in all-ingoing convention.\label{fig:TNjk-2-def}}
\end{figure}
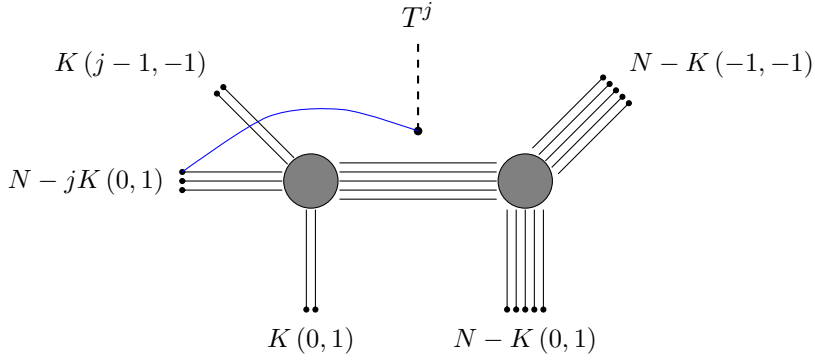

Gauge theory deformations of these theories can be read off from the mass deformation shown in fig.~\ref{fig:TNjk-2-def}, where those D7-branes which have multiple D5-branes attached were moved to a position where they have no D5-branes attached. 
For $N>jK$ the quiver gauge theory is
\begin{align}\label{eq:TN-quiver}
[N-jK]\stackrel{y_1}{-}(N-jK+j-1)\stackrel{x_1}{-}
\dots\stackrel{x_{K-1}}{-}(N&-K)\stackrel{x_K}{-}
\cdots \stackrel{x_{N-3}}{-}(2)\stackrel{y_3}{-}[2]\nonumber \\
&\ \,| \, {\scriptstyle y_2}\\
&\,[j]\nonumber
\end{align}
where $(N)$ denotes an $SU(N)$  gauge node and $[M]$ denotes $M$ hypermultiplets in the fundamental representation of the gauge group node they are attached to. 
Between the links labeled by $x_1$ and $x_{K-1}$ are $K-2$ gauge nodes, with the gauge group rank increasing in steps of $j-1$. 
For $j=1$ this just gives a total of $K$ $SU(N-K)$ gauge nodes. Between the links labeled by $x_{K}$ and $x_{N-3}$ the gauge group rank decreases in steps of one.
Associated with each link between gauge group nodes is a hypermultiplet in the bifundamental representation.
We denote by $(x_i,\tilde x_i)$ the scalars in the bifundamental hypermultiplets, and by $(y_i,\tilde y_i)$ the scalars in the fundamental hypermultiplets.

There is a meson operator connecting the $[N-jK]$ fundamentals on the left end of the quiver to the $j$ fundamental hypermultiplets at the gauge $SU(N-K)$ node,
\begin{align}\label{eq:T-Nkj-op}
 \mathcal O_\alpha&= (y_1\cdot x_1\cdots x_{K-1}\cdot y_2)_\alpha^{\tilde\beta}~,
\end{align}
where $\alpha$ is an $SU(N-K)$ index and $\tilde\beta$ an $SU(j)$ index. 
The scaling dimension is
\begin{align}\label{eq:T-Njk-Delta}
 \Delta(\mathcal O_\alpha)&=\frac{3}{2}(K+1)~.
\end{align}
This operator is in the $({\bf N-jK},{\bf \bar j})$ representation of the $SU(N-jK)\times SU(j)$ part of the global symmetry. In the 5-brane junction it is realized by a fundamental string connecting D7-branes and D5-branes as shown in fig.~\ref{fig:TNjk-2-def}. There are further meson operators in the field theory which correspond to string junctions.

For $jK=N$ there are no unconstrained D5-branes. 
For $j=2$ and $N=2K$, the configuration in fig.~\ref{fig:TNjk-2-def} is a gluing of two unconstrained $T_{K}$ theories, gauging their $SU(K)$ flavor symmetries. In the gauge theory deformation the D7 branes add flavor to the gauged $SU(K)$ node, and the quiver gauge theory is
\begin{align}
 [2]-(2)-(3)-\cdots - (K-1)- (&K) - (K-1) -\cdots -(3)-(2)-[2] \nonumber\\
 &\,| \\
 &\,\!\![2]\nonumber
\end{align}
For $N=4$, $K=j=2$, this is $SU(2)$ with six flavors, realizing the $E_7$ theory. 
For $j>2$ and $N=jK$ the gauge theory deformation is
\begin{align}
 (j-1)-(2(j-1))-\cdots  - (N&-K) - (N-K-1) -\cdots -(2)-[2] \nonumber\\
 &\, \ | \\
 &\, \ \!\![2]\nonumber
\end{align}
The meson operators in these theories correspond to string junctions.

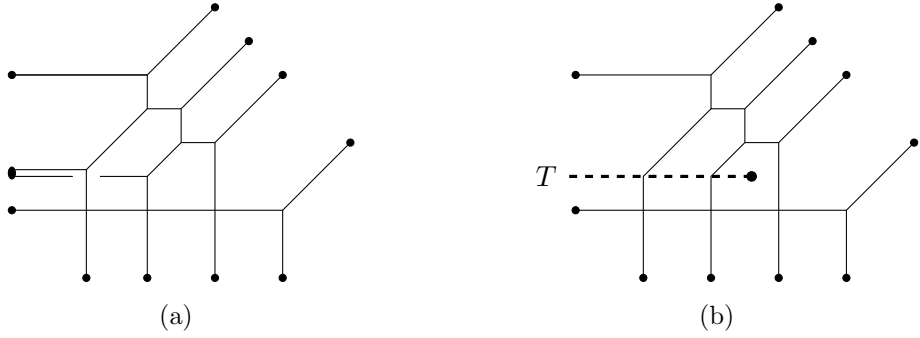
\begin{figure}
 \centering
 \subfigure[][]{
  \begin{tikzpicture}[scale=0.9]
    \draw (-2,0) -- (2,0) -- (2,-1);
    \draw (2,0) -- (3,1);
    
    \draw (1,-1) -- (1,1) -- (2,2);
    \draw (1.0,1) -- (0.5,1) -- (0,0.5) -- (0,-1);
    \draw (0.5,1) -- (0.5,1.5) -- (1.5,2.5);
    \draw (0.5,1.5) -- (0,1.5) -- (0,2) -- (1,3);
    \draw (0,1.5) -- (-0.9,0.6) -- (-0.9,-1);
    \draw (0,2) -- (-2,2);
    
    \draw (-0.9,0.6) -- (-2,0.6);
    \draw (0.0,0.5) -- (-0.7,0.5);
    \draw (-1.1,0.5) -- (-2,0.5);
    
    \draw (0,2) -- (-2,2);
    
    \foreach \i in {(-0.9,-1),(0,-1),(1,-1),(2,-1),(-2,0),(3,1),(2,2),(1.5,2.5),(1,3),(-2,2)}{
     \draw[fill] \i circle (1.5pt);
    }
    \draw[fill] (-2,0.55) ellipse (1.5pt and 2.5pt);
  \end{tikzpicture}
 }\hskip 20mm
 \subfigure[][]{
  \begin{tikzpicture}[scale=0.9]
    \draw (-2,0) -- (2,0) -- (2,-1);
    \draw (2,0) -- (3,1);
    \draw[fill] (0.6,0.5) circle (2pt);
    \draw[very thick,dashed] (0.6,0.5) -- (-2.1,0.5) node [anchor=east] {\small $T$};
    
    \draw (1,-1) -- (1,1) -- (2,2);
    \draw (1.0,1) -- (0.5,1) -- (0,0.5) -- (0,-1);
    \draw (0.5,1) -- (0.5,1.5) -- (1.5,2.5);
    \draw (0.5,1.5) -- (0,1.5) -- (0,2) -- (1,3);
    \draw (0,1.5) -- (-1.0,0.5) -- (-1.0,-1);
    \draw (0,2) -- (-2,2);
    
    \foreach \i in {(-1,-1),(0,-1),(1,-1),(2,-1),(-2,0),(3,1),(2,2),(1.5,2.5),(1,3),(-2,2)}{
     \draw[fill] \i circle (1.5pt);
    }
  \end{tikzpicture}
 }
 \caption{5-brane webs for the $T_{N,K,j}$ theory with $N=4$, $K=2$, $j=1$. On the left hand side as a constrained triple junction, on the right hand side with the 7-branes inside the web.\label{fig:TN-web-open}}
\end{figure}

\subsection{The \texorpdfstring{$T_{N,K,j}$}{T-N-K-j} solution}

\begin{figure}
\centering
\begin{tikzpicture}[scale=0.9]
\draw[fill=lightgray,opacity=0.2] (0,0) circle (1.5);
\draw[thick] (0,0) circle (1.5);
\draw[thick] ({1.4/sqrt(2)},{1.4/sqrt(2)}) -- ({1.6/sqrt(2)},{1.6/sqrt(2)});
\draw[thick] (0,-1.4) -- (0,-1.6);
\draw[thick] (-1.4,0) -- (-1.6,0);
\node at (-2.85,0) {\small $N-jK$ D5};
\node at (0.2,1.0) {$\mathbf \Sigma$};
\node at (0,-1.92) {\small $N$ NS5};
\node at (1.65,1.46) {\small $N$ $(1,1)$};

\draw[fill=black] (-0.3,0) circle (0.05);
\draw[thick,dashed, black] (-0.3,0) node [anchor=north] {\small D7} -- (-1.5,0);
\node at (-0.8,0) [anchor=south] {\small $T^j$};
\end{tikzpicture}
\caption{Disc representation of the supergravity solution corresponding to the $T_{N,K,j}$ junction after moving the constrained groups of 5-branes into the web.\label{fig:TN-disc}}
\end{figure}
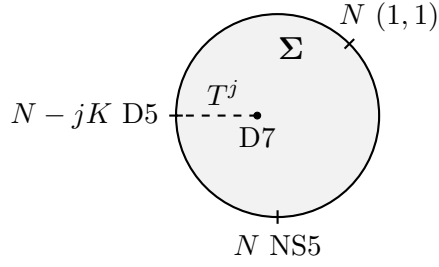

Supergravity duals for the $T_{N,K,j}$ theories can be realized by considering the brane web for a general mass deformation of the SCFT and moving the D7 branes with multiple D5-branes ending on them into the web, as shown in fig.~\ref{fig:TN-web-open}. The supergravity solution corresponding to the conformal limit of the resulting brane web then takes the form shown in fig.~\ref{fig:TN-disc}, with three poles on the boundary of $\Sigma$, corresponding to unconstrained D5, NS5 and $(1,1)$ 5-branes, as well as a puncture with D7-brane monodromy.
For $N=jK$, the D5-brane charge at the pole vanishes and the solution reduces to a two-pole solution.

For the explicit realization $\Sigma$ is taken as the upper half plane and the general ansatz has three poles. The $SL(2,\RR)$ automorphisms can be used to fix
\begin{align}
 r_1&=1~, & r_2&=0~, & r_3&=-1~.
\end{align}
This exhausts the freedom to make $SL(2,\RR)$ transformations and the punctures can not from the outset be fixed to the imaginary axis. The residues corresponding via (\ref{eq:cY-pq}) to the 5-brane charges in fig.~\ref{fig:TN-disc} are
\begin{align}\label{eq:T-Njk-cY}
 \cY_+^1&=\frac{3}{4}\alpha^\prime N~,
 &
 \cY_+^2&=\frac{3}{4}\alpha^\prime i(N-jK)~,
 &
 \cY_+^3&=-\frac{3}{4}\alpha^\prime (1+i)N~.
\end{align}
For $jK=N$ the residue at $r_2$ vanishes and the solution reduces to a two-pole solution.
The appropriate monodromy for the puncture is realized by fixing in (\ref{eq:f-def})
\begin{align}
 I&=1~, & n_1^2&=j~, & \gamma_1&=1~.
\end{align}
All relevant brane charges, $N$, $K$ and $j$, are manifest in the supergravity solution.
The parameters realizing the residues (\ref{eq:T-Njk-cY}), via (\ref{eqn:residues}) and (\ref{eq:cY}), and solving the regularity conditions (\ref{eq:reg-gen}) are
\begin{align}
 \sigma&=\frac{3}{4}\alpha^\prime(2+i)N~, & s_1&=\frac{\cY_+^2}{\sigma}~, &
 w_1&=i\tan\frac{\pi K}{2N}~,
\end{align}
with $\cA_0^+=\frac{1}{2}J_1+\cY_+^2\ln 2$. 
As $jK$ approaches $N$ from below, the zero $s_1$ ($\bar s_1$) of $\partial_w\cA_+$ ($\partial_w\cA_-$) approaches the pole at $r_2$, annihilating it for $N=jK$, such that the configuration reduces to a two-pole solution.
The puncture is fixed to the imaginary axis by the regularity conditions, and the branch cut intersects the pole at $r_2$.  As discussed in \cite{Gutperle:2018vdd} this does not affect the regularity conditions for poles with imaginary residue.

\subsubsection{String embedding}

Holographically, the meson operator (\ref{eq:T-Nkj-op}) is realized as a string embedded into the solution.
The equation of motion for $(p,q)$ strings embedded along the time coordinate of global $AdS_6$ and a one-dimensional curve in $\Sigma$, as well as the expressions for the scaling dimension and $R$-charge of the dual operator, were derived in sec.~4.1 of \cite{Bergman:2018hin}. The equation of motion for the embedding function $w(\xi)$ reads 
\begin{align}\label{eq:pq-eom}
 0&=\frac{\bar w^{\prime\prime}}{\bar w^\prime}-\frac{w^{\prime\prime}}{w^\prime}+\left(\bar w^\prime\partial_{\bar w}-w^\prime\partial_w\right)\ln\big(f_6^2\rho^2 \q^T\cM\q\big)~,
\end{align}
where $\q^T\cM \q=e^{2\phi}(p-q\chi)^2+e^{-2\phi}q^2$.
Scaling dimension and charge are given by
\begin{align}\label{eq:pq-Delta}
 \Delta_{(p,q)}&=-2T\int_{\Sigma_{(p,q)}} d\xi f_6 \rho|w^\prime|\sqrt{\q^T\cM\q}~,\\
  Q_{(p,q)}&=T\int_{\Sigma_{(p,q)}} \left(p\Re(d\cC)-q\Im(d\cC)\right)~.
  \label{eq:pq-Q}
\end{align}

For the three-pole solutions corresponding to the $T_{N,K,j}$ junctions, we expect to find a fundamental string connecting one of the $j$ D7-branes on which $K$ constrained 5-branes end to one of the unconstrained D5-branes (fig.~\ref{fig:TNjk-2-def}). This string naturally transforms in the $({\bf N-jK},{\bf \bar j})$ representation of the $SU(N-jK)\times SU(j)$ part of the global symmetry (\ref{eq:TN-symmetry}).
We indeed find a fundamental string embedded into the solution, connecting the D7-brane puncture and the D5-brane pole along the imaginary axis in the upper half plane. 
This embedding lies on the branch cut, but the terms in the equation of motion and the integrands in (\ref{eq:pq-Delta}), (\ref{eq:pq-Q}) are single-valued for $(p,q)=(1,0)$.
The embedding solves the equation of motion and yields
\begin{align}
 \Delta_{\rm F1}&=\frac{3}{2}K~, & Q_{\rm F1}&=\frac{1}{2}K~.
\end{align}
The scaling dimension is related to the charge by the expected BPS relation and both agree with the field theory result (\ref{eq:T-Njk-Delta}) at large $K$.

\subsection{The \texorpdfstring{$+_{N,M,k,j}$}{plustilde} theory}

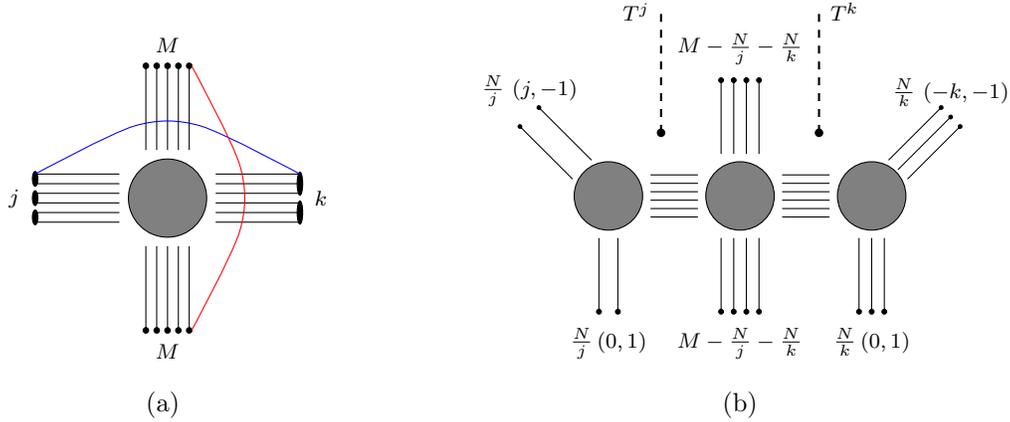
\begin{figure}
\centering
\subfigure[][]{\label{fig:D5NS5-3}
  \begin{tikzpicture}[scale=0.8]
  \draw[fill=gray] (0,0) circle (0.65);
  \foreach \i in {-5,-3,-1,1,3,5}{
   \draw (0.8,0-0.08*\i) -- +(1.4,0) [fill];
   \draw (-0.8,0-0.08*\i) -- +(-1.4,0) [fill];
  }
  \draw[fill] (0.8+1.4,0.08*3) ellipse (1.3pt and 5.5pt);
  \draw[fill] (0.8+1.4,-0.08*3) ellipse (1.3pt and 5.5pt);
  
  \draw[fill] (-0.8-1.4,0.08*4) ellipse (1.3pt and 3.5pt);
  \draw[fill] (-0.8-1.4,0) ellipse (1.3pt and 3.5pt);
  \draw[fill] (-0.8-1.4,-0.08*4) ellipse (1.3pt and 3.5pt);

  \foreach \i in {-6,-3,0,3,6}{
   \draw (0.06*\i,0.8) -- +(0,1.4) [fill] circle (1.3pt);
   \draw (0.06*\i,-0.8) -- +(0,-1.4) [fill] circle (1.3pt);
  }
  \node at (0.8+1.75,0) {\scriptsize $k$};
  \node at (-0.8-1.75,0) {\scriptsize $j$};
  \node at (0,0.8+1.75) {\scriptsize $M$};
  \node at (0,-0.8-1.75) {\scriptsize $M$};
  
  \draw[blue] plot [smooth] coordinates { (-2.2,0.4) (-0.55,1.2) (0.55,1.2)  (2.2,0.4)};
  \draw[red] plot [smooth] coordinates { (0.4,-2.2) (1.2,-0.55) (1.2,0.55)  (0.4,2.2)};
  \end{tikzpicture}
}\hskip 15mm
\subfigure[][]{\label{fig:D5NS5-def}
\begin{tikzpicture}[scale=0.7]
  \draw[fill=gray] (0,0) circle (0.65);
  \draw[fill=gray] (2.5,0) circle (0.65);
  \draw[fill=gray] (-2.5,0) circle (0.65);
  
  \foreach \i in {-6,-2,2,6}{
   \draw (0.06*\i,0.8) -- +(0,1.4) [fill] circle (1.3pt);
   \draw (0.06*\i,-0.8) -- +(0,-1.4) [fill] circle (1.3pt);
  }
  \foreach \i in {-5,-3,-1,1,3,5}{
   \draw (0.8,0-0.08*\i) -- +(0.9,0);
   \draw (-0.8,0-0.08*\i) -- +(-0.9,0);
  }
  \foreach \i in {-3,3}{
   \draw (-2.5+0.06*\i,-0.8) -- +(0,-1.4) [fill] circle (1.3pt);
  }
  \foreach \i in {-4,0,4}{
   \draw (2.5+0.06*\i,-0.8) -- +(0,-1.4) [fill] circle (1.3pt);
  }
  \foreach \i in {-2,0,2}{
    \draw (2.5+0.5+0.09*\i,0.5-0.09*\i) -- (2.5+1.5+0.09*\i,1.5-0.09*\i) [fill=black] circle (1pt);
  }
  \foreach \i in {-2,2}{
    \draw (-2.5-0.5-0.09*\i,0.5-0.09*\i) -- (-2.5-1.5-0.09*\i,1.5-0.09*\i) [fill=black] circle (1pt);
  }
  \node [anchor=south] at (0,2.3) {\scriptsize $M-\frac{N}{j}-\frac{N}{k}$};
  \node [anchor=north] at (0,-2.3) {\scriptsize $M-\frac{N}{j}-\frac{N}{k}$};
  \node [anchor=north] at (-2.5,-2.3) {\scriptsize $\frac{N}{j}\,(0,1)$};
  \node [anchor=north] at (2.5,-2.3) {\scriptsize $\frac{N}{k}\,(0,1)$};
  \node [anchor=south] at (4,1.5) {\scriptsize $\frac{N}{k}$ $(-k,-1)$};
  \node [anchor=south] at (-4,1.5) {\scriptsize $\frac{N}{j}$ $(j,-1)$};
  
  \draw[thick,dashed] (1.5,1.2) -- (1.5,3.5) node [anchor=west] {\scriptsize $T^k$};
  \draw[fill] (1.5,1.2) circle (2pt);
  
  \draw[thick,dashed] (-1.5,1.2) -- (-1.5,3.5) node [anchor=east] {\scriptsize $T^j$};
  \draw[fill] (-1.5,1.2) circle (2pt);
\end{tikzpicture}
}
\caption{The $+_{N,M,k,j}$ theory with $N=6$, $M=5$, $j=3$ and $k=2$ in \ref{fig:D5NS5-3}, along with a D1 (red vertical curve) and an F1 string (blue horizontal curve)  representing the stringy operators. 
Fig.~\ref{fig:D5NS5-def} shows a mass deformation for $M\geq \frac{N}{j}+\frac{N}{k}$.\label{fig:plus-NMkj}}
\end{figure}

The second class of 5d SCFTs is realized by quartic junctions, constructed from two groups of $M$ unconstrained NS5 branes and two groups of $N$ D5-branes. One group of $N$ D5-branes is partitioned into $k>N/M$ subgroups of $N/k$ that end on the same D7-brane, the other group of $N$ D5-branes is partitioned into $j>N/M$ subgroups of $N/j$ that end on the same D7-brane, fig.~\ref{fig:D5NS5-3}. 
The $+_{N,M,k,j}$ and $+_{N,M,j,k}$ theories are related by a parity transformation. 
For $k=j=N$ the $+_{N,M,k,j}$ junctions include the unconstrained D5/NS5 intersection discussed initially in \cite{Aharony:1997bh}, and for $j=N$ they include the $+_{N,M,k}$ theories of \cite{Bergman:2018hin}.
The global symmetry of the $+_{N,M,k,j}$ SCFT is 
\begin{align}
 SU(M)^2\times SU(k)\times SU(j)\times U(1)~.
\end{align}
We expect stringy operators in the $({\bf 1},{\bf 1},{\bf k},{\bf \bar j})$ and $({\bf M},{\bf \bar M},{\bf 1},{\bf 1})$ representations of this group, realized as F1 and D1 strings as shown in fig.~\ref{fig:D5NS5-3}.

For $k=1$ (or $j=1$), the $+_{N,M,k,j}$ junction is equivalent to a $T_N$ theory. This can be seen for $k=1$ by moving the single D7-brane on the right hand side in fig.~\ref{fig:D5NS5-3} all the way to the left. The attached $N$ D5 branes are converted to $M-N$ D5-branes by Hanany-Witten transitions, and the $M$ NS5-branes pointing upwards are converted to $(1,1)$ 5-branes. The result is a junction of $M$ unconstrained NS5 branes with $M$ unconstrained $(1,1)$ 5-branes and $M$ D5-branes partitioned into groups ending on the same D7-brane as $Y_1=[(N/j)^j,M-N]$.
In particular, the $+_{2,4,1,1}$ junction gives an alternative realization of the $E_7$ theory
and the $+_{2,3,1,2}$ junction gives an alternative realization of the $E_6$ theory.
Moreover, for $M=5$ and $k=j=1$ with either $N=2$ or $N=3$ the $+_{N,M,k,j}$ junction realizes the $S_5$ theory of \cite{Bergman:2014kza}.

For $j=k$, moving the $k$ D7-branes on the right all the way to the left and the $j$ D7-branes on the left all the way to the right leads to another intersection of the $+_{N,M,k,j}$ form. 
We find the equivalence  $+_{N,M,k,k}=+_{kM-N,M,k,k}$.
In particular, for $j=k=N/(M-1)$ the $+_{N,M,k,j}$ junctions are equivalent to unconstrained junctions.
The $+_{N,N+1,1,1}$ junctions are equivalent to $+_{1,N+1,1,1}$ junctions, which have an S-dual description as $(N+1)^2$ free hypermultiplets.

For the discussion of more general gauge theory deformations we restrict the parameters to $M\geq\frac{N}{k}+\frac{N}{j}$ (in addition to $k,j>N/M$). The case of smaller $M$ can be treated analogously. The gauge theory can be read off after mass deforming and moving the D7-branes as in fig.~\ref{fig:D5NS5-def}. The quiver gauge theory is given by
\begin{align}\label{eq:plus-quiver}
(j)\stackrel{x_1}{-}(2j)\stackrel{x_2}{-}\cdots \stackrel{x_{\frac{N}{j}-1}}{-}(&N)\stackrel{x_\frac{N}{j}}{-}(N)^{M - \frac{N}{k}-\frac{N}{j}-1}\stackrel{x_{M-\frac{N}{k}-1}}{-}(N)\stackrel{x_{M-\frac{N}{k}}}{-}\cdots 
\stackrel{x_{M-3}}{-} (2k)\stackrel{x_{M-2}}{-}(k) \nonumber \\
&\:| \, {\scriptstyle y_1}\hspace*{50mm}| \, {\scriptstyle y_2}\\
&\![j] \hspace*{51mm} [k]\nonumber
\end{align}
For $M=\frac{N}{k}+\frac{N}{j}$ there is one $SU(N)$ gauge node in the center with $[k+j]$ flavors. For $k=1$ the would-be $(1)$ on the right end is replaced by two fundamental hypermultiplets, $[2]$, and likewise on the left end for $j=1$.

The $SU(k)\times SU(j)$ part of the SCFT global symmetry is manifest in the gauge theory as flavor symmetry. 
The stringy operators in the gauge theory include the mesons
\begin{align}\label{eq:meson-plus}
 \mathcal O_\alpha^{\tilde\beta}&= (y_1\cdot x_{N/j}\cdots x_{M-1-N/k}\cdot y_2)_\alpha^{\tilde\beta}~,
\end{align}
where $\alpha$ is an $SU(j)$ index and $\tilde \beta$ an $SU(k)$ index. This operator is in the $({\bf k},{\bf \bar j})$ representation of $SU(k)\times SU(j)$ and the scaling dimension is 
\begin{align}\label{eq:Delta-meson-plus}
 \Delta(\mathcal O_\alpha^{\tilde\beta})&=\frac{3}{2}\left(M+2-\frac{N}{j}-\frac{N}{k}\right)~. 
\end{align}
Moreover, there are $SU(k)\times SU(j)$ singlet operators with scaling dimension $3/2N$,
\begin{align}\label{eq:plus-D1-op}
 \mathcal O&=\det(x_i)~, & i&=\frac{N}{j},\ldots,M-1-\frac{N}{k}~,
\end{align}
where the $x_i$ are bifundamentals between $SU(N)$ gauge nodes and we introduced the short hand notation $\det(x_i)=\epsilon^{\alpha_1\ldots \alpha_N}\epsilon_{\beta_1\ldots \beta_N}(x_i)_{\alpha_1}^{\beta_1}\ldots (x_i)_{\alpha_N}^{\beta_N}$. These are components of the $({\bf M},{\bf \bar M},{\bf 1},{\bf 1})$ SCFT operator.
In the 5-brane junction the meson operator (\ref{eq:meson-plus}) is expected to be realized by a fundamental string, and the operators in (\ref{eq:plus-D1-op}) by a D1-brane, as shown in fig.~\ref{fig:D5NS5-3}.

\subsection{The \texorpdfstring{$+_{N,M,k,j}$}{plustilde} solution}

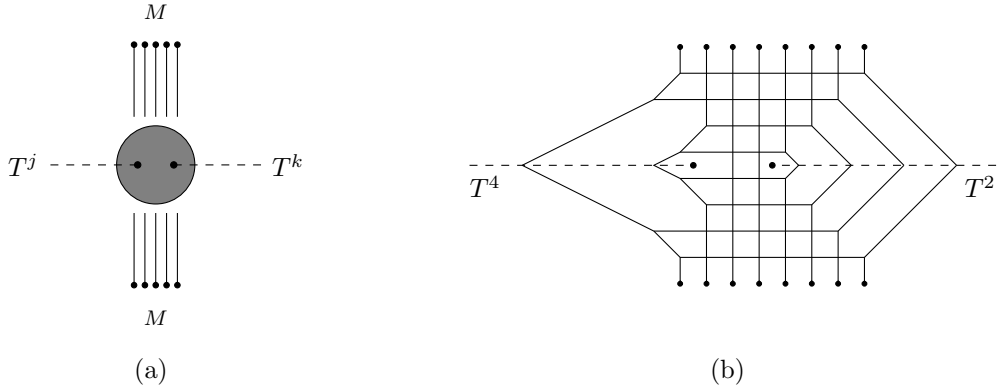
\begin{figure}
\centering
\subfigure[][]{\label{fig:plus-intersection}
  \begin{tikzpicture}[scale=0.8]
  \draw[fill=gray] (0,0) circle (0.65);
  
  \foreach \i in {-6,-3,0,3,6}{
   \draw (0.06*\i,0.8) -- +(0,1.2) [fill] circle (1.3pt);
   \draw (0.06*\i,-0.8) -- +(0,-1.2) [fill] circle (1.3pt);
  }
  
  \draw[fill] (0.3,0) circle (1.5pt);
  \draw[dashed] (0.3,0) -- +(1.5,0);
  \node at (2.2,0) {\footnotesize $T^k$};
  \draw[fill] (-0.3,0) circle (1.5pt);
  \draw[dashed] (-0.3,0) -- +(-1.5,0);
  \node at (-2.2,0) {\footnotesize $T^j$};
  
  \node at (0,0.8+1.75) {\scriptsize $M$};
  \node at (0,-0.8-1.75) {\scriptsize $M$};
  \end{tikzpicture}
 }\qquad\qquad
\subfigure[][]{
   \begin{tikzpicture}[scale=0.7]
      \draw[dashed] (-5,0) -- (-0.75,0);
      \node at (-4.7,-0.4) {\footnotesize $T^4$};
      \draw[fill] (-0.75,0) circle (1.5pt);
      \draw[dashed] (5,0) -- (0.75,0);
      \node at (4.7,-0.4) {\footnotesize $T^2$};
      \draw[fill]  (0.75,0) circle (1.5pt) ;
      
      \draw (1.25,0) -- +(-0.25,0.25) -- (-1,0.25) -- (-1.5,0) -- (-1,-0.25) -- (1.0,-0.25) -- (1.25,0);
      \draw (2.25,0) -- +(-0.75,0.75) -- (-0.5,0.75) -- (-1,0.25);
      \draw (2.25,0) -- +(-0.75,-0.75) -- (-0.5,-0.75) -- (-1,-0.25);
      \draw (3.25,0) -- +(-1.25,1.25) -- (-1.5,1.25);
      \draw (3.25,0) -- +(-1.25,-1.25) -- (-1.5,-1.25);
      \draw (4.25,0) -- +(-1.75,1.75) -- (-1.0,1.75) -- (-1.5,1.25) -- (-4.0,0);
      \draw (4.25,0) -- +(-1.75,-1.75) -- (-1,-1.75) -- (-1.5,-1.25) -- (-4.0,0);

      \draw[fill] (2.5,1.75) -- +(0,0.5) circle (1.3pt);
      \draw[fill] (2.5,-1.75) -- +(0,-0.5) circle (1.3pt);
      
      \draw[fill] (2.0,1.25) -- +(0,1.0) circle (1.3pt);
      \draw[fill] (2.0,-1.25) -- +(0,-1.0) circle (1.3pt);

      \draw[fill] (1.5,0.75) -- +(0,1.5)  circle (1.3pt);
      \draw[fill] (1.5,-0.75) -- +(0,-1.5) circle (1.3pt);
      \draw[fill] (1,0.25) -- +(0,2) circle (1.3pt);
      \draw[fill] (1,-0.25) -- +(0,-2) circle (1.3pt);
      \draw[fill] (-0.5,0.75) -- +(0,1.5) circle (1.3pt);
      \draw[fill] (-0.5,-0.75) -- +(0,-1.5) circle (1.3pt);
      \draw[fill] (-1.0,1.75) -- +(0,0.5) circle (1.3pt);
      \draw[fill] (-1.0,-1.75) -- +(0,-0.5) circle (1.3pt);
      
      \draw[fill] (0,2.25) circle (1.3pt) -- (0,-2.25) circle (1.3pt);
      \draw[fill] (0.5,2.25) circle (1.3pt) -- (0.5,-2.25) circle (1.3pt);
      
      \draw[fill,white] (0,-3.16) circle (2pt);
   \end{tikzpicture}
 }
 \caption{The $+_{N,M,k,j}$ 5-brane junction with the 7-branes moved into the intersection on the left hand side. On the right a gauge theory deformation with $N=M=8$, $j=4$, $k=2$.\label{fig:plus-web}}
\end{figure}

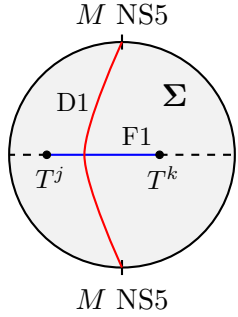
\begin{figure}
\centering
\begin{tikzpicture}
\draw[fill=lightgray,opacity=0.2] (0,0) circle (1.5);
\draw[thick] (0,0) circle (1.5);
\draw[thick] (0,1.4) -- (0,1.6);
\draw[thick] (0,-1.4) -- (0,-1.6);
\node at (0.7,0.8) {$\mathbf \Sigma$};
\node at (0,1.87) {\small $M$ NS5};
\node at (0,-1.92) {\small $M$ NS5};

\draw[thick,blue] (-1,0) -- (0.5,0);
\draw[thick, red] plot [smooth] coordinates { (0,1.5) (-0.5,0)  (0,-1.5)};

\draw[fill=black] (0.5,0) circle (0.05);
\draw[fill=black] (-1.0,0) circle (0.05);
\draw[thick,dashed, black] (0.5,0) -- (1.5,0);
\draw[thick,dashed, black] (-1.0,0) -- (-1.5,0);
\node at (0.55,-0.3) {\footnotesize $T^k$};
\node at (-0.95,-0.3) {\footnotesize $T^j$};
\node at (0.2,0.25) {\footnotesize F1};
\node at (-0.65,0.7) {\footnotesize D1};
\end{tikzpicture}
\caption{Disc representation of the $+_{N,M,k,j}$ solution. The straight horizontal blue line represents a fundamental string stretching between the D7-brane punctures, the solid red curve represents a D1 brane stretching between the NS5-brane poles.\label{fig:plus-disc}}
\end{figure}

The supergravity solution corresponding to the junction in fig.~\ref{fig:D5NS5-3} is constructed by moving the D7-branes into the web, as shown in fig.~\ref{fig:plus-web}, and given by a two-pole solution with two D7-brane punctures, as illustrated in fig.~\ref{fig:plus-disc}. Unlike for the solutions discussed so far, the brane charges in the solution alone do not uniquely identify the associated 5-brane junction. The number of NS5-branes, $M$, and the numbers of D7-branes, $k$ and $j$, are manifest in the solution as residues and monodromies, but the total number of D5-branes, $N$, is not. The location of the punctures, however, is expected to be different for different $N$. We will use one of the stringy operators discussed above to precisely identify the supergravity solution associated with a given 5-brane junction.

The two punctures in the supergravity solutions corresponding to the $+_{N,M,k,j}$ theories are realized by fixing $I=2$ in (\ref{eq:f-def}) and
\begin{align}
 \alpha_1&<\alpha_2~, &\gamma_1&=-\gamma_2=1~, & n_1^2&=j~, & n_2^2&=k~.
\end{align}
The branch cuts do not intersect for $\alpha_1<\alpha_2$.
With this choice of branch cut orientations, $f(x)=-f(-x)$ for $x\in\RR$. 
A convenient parametrization for the locations of the punctures is
\begin{align}
 \alpha_1&=\tan\theta_1~, & \alpha_2&=\cot\theta_2~.
\end{align}
Without loss of generality we can choose $0<\theta_1,\theta_2<\pi/2$. 
The residues realizing the NS5-brane poles via (\ref{eq:cY-pq}) are
\begin{align}
 \cY_+^{1}=-\cY_+^2&=\frac{3}{4}\alpha^\prime M~.
\end{align}
The first equality implies  $f(1)=f(-1)=0$, and these residues are realized for
\begin{align}\label{eq:jsol}
 k\theta_2&=j \theta_1~,
 &
 Z_+&=\frac{3}{4}\alpha^\prime M~.
\end{align}
This fixes one of $\theta_1$, $\theta_2$. In addition to the 5-brane junction parameters $M$, $k$, $j$, which are realized directly in the supergravity solution as poles and punctures, the solution has one real degree of freedom in $\theta_1$, $\theta_2$, which encodes $N$. The precise relation will be determined next.

\subsubsection{String embeddings}

The string theory realization of the $({\bf 1},{\bf 1},{\bf k},{\bf \bar j})$ and $({\bf M},{\bf \bar M},{\bf 1},{\bf 1})$ operators as F1 and D1 strings, respectively, as shown in fig.~\ref{fig:D5NS5-3}, leads to the supergravity representation shown in fig.~\ref{fig:plus-disc}.
The brane intersection in fig.~\ref{fig:plus-intersection} has a $\ZZ_2$ symmetry acting as reflection across the horizontal line along which the branch cuts are oriented. In the upper half plane with poles at $r_1=-r_2=1$, this symmetry acts as reflection across the imaginary axis. 
An embedding of a fundamental string along a straight line connecting the two punctures on the imaginary axis consequently solves the equations of motion. Scaling dimension and charge of the string are
\begin{align}\label{eq:Delta-F1-plus}
 \Delta_{\rm F1}&=3Q_{\rm F1}~,
 &
 Q_{\rm F1}&=\left(\frac{1}{2}-\frac{\theta_1+\theta_2}{\pi}\right)M~.
\end{align}
This also satisfies the expected BPS relation. In the limit $\theta_1,\theta_2\rightarrow 0$, in which the punctures approach the boundary, the scaling dimension reduces to the result for the F1 operator in the $+_{N,M}$ theory \cite{Bergman:2018hin}, as expected.

The embedding of the D1 brane is less straightforward to find, but the scaling dimension can be determined as follows. For $k=j$ there is a second $\ZZ_2$ symmetry, which on the disc as in fig.~\ref{fig:plus-disc} acts as reflection across the vertical diameter. 
Embedding the D1 along the straight line connecting the poles consequently solves the equations of motion for $j=k$. Deforming the solution to $k\neq j$ will deform the embedding, in general to a curve as shown in fig.~\ref{fig:plus-disc}. The $R$-charge (\ref{eq:pq-Q}), however, can be determined from the end points of the D1-brane alone. This yields
\begin{align}\label{eq:D1-op-plus}
 Q_{\rm D1}&=\frac{M k\theta_2}{\pi} ~.
\end{align}
The scaling dimension is then fixed by the BPS relation, to $\Delta_{\rm D1}=3Q_{\rm D1}$.

Since the D1 is expected to realize the operator (\ref{eq:plus-D1-op}), with $\Delta=3/2N$, this result can be used to solve for $N$ in terms of $\theta_1$ or $\theta_2$, which yields
\begin{align}\label{eq:N-plus}
 N&=\frac{2Mk \theta_2}{\pi}=\frac{2Mj\theta_1}{\pi}~.
\end{align}
This completes the identification of parameters in the supergravity solution with those of the 5-brane junction.
With the expression for $N$ in (\ref{eq:N-plus}), the scaling dimension of the $F1$ operator in (\ref{eq:Delta-F1-plus}) becomes
\begin{align}
 \Delta&=\frac{3}{2}\left(M-\frac{N}{k}-\frac{N}{j}\right)~.
\end{align}
This agrees with the field theory result (\ref{eq:Delta-meson-plus}) at large $N$, $M$, providing a non-trivial consistency check and confirming the interpretation of the supergravity solutions as holographic duals for the $+_{N,M,k,j}$ theories.
The restriction $\alpha_1<\alpha_2$, which ensures that the branch cuts do not intersect in the supergravity solution, translates to $\frac{N}{j}+\frac{N}{k}<M$. 
This is the assumption used to derive the quiver (\ref{eq:plus-quiver}). 
Supergravity solutions for smaller values of $M$ can be constructed by first rotating the branch cuts in the brane intersection in fig.~\ref{fig:plus-intersection} and then identifying the corresponding two-pole solution.

\begin{acknowledgments}

We thank Oren Bergman for interesting discussions. 
Part of this work was completed at the Aspen Center for Physics, which is supported by National Science Foundation grant PHY-1607611. We thank the organizers and participants of the program ``Superconformal Field Theories and Geometry'' for the interesting workshop.
CFU thanks the organizers and participants of the workshop ``Strings, Branes and Gauge Theories'' for the interesting workshop and APCTP for hospitality and support.
This work is supported in part by the National Science Foundation under grant PHY-16-19926.  
\end{acknowledgments}

\bibliographystyle{JHEP}
\bibliography{2pole}

\providecommand{\href}[2]{#2}\begingroup\raggedright\begin{thebibliography}{10}

\bibitem{Seiberg:1996bd}
N.~Seiberg, \emph{{Five-dimensional SUSY field theories, nontrivial fixed
  points and string dynamics}},
  \href{http://dx.doi.org/10.1016/S0370-2693(96)01215-4}{\emph{Phys. Lett.}
  {\bf B388} (1996) 753--760}, [\href{http://arxiv.org/abs/hep-th/9608111}{{\tt
  hep-th/9608111}}].

\bibitem{Morrison:1996xf}
D.~R. Morrison and N.~Seiberg, \emph{{Extremal transitions and five-dimensional
  supersymmetric field theories}}, {\emph{Nucl. Phys.} {\bf B483} (1997)
  229--247}, [\href{http://arxiv.org/abs/hep-th/9609070}{{\tt
  hep-th/9609070}}].

\bibitem{Douglas:1996xp}
M.~R. Douglas, S.~H. Katz and C.~Vafa, \emph{{Small instantons, Del Pezzo
  surfaces and type I-prime theory}},
  \href{http://dx.doi.org/10.1016/S0550-3213(97)00281-2}{\emph{Nucl. Phys.}
  {\bf B497} (1997) 155--172}, [\href{http://arxiv.org/abs/hep-th/9609071}{{\tt
  hep-th/9609071}}].

\bibitem{Intriligator:1997pq}
K.~A. Intriligator, D.~R. Morrison and N.~Seiberg, \emph{{Five-dimensional
  supersymmetric gauge theories and degenerations of Calabi-Yau spaces}},
  \href{http://dx.doi.org/10.1016/S0550-3213(97)00279-4}{\emph{Nucl. Phys.}
  {\bf B497} (1997) 56--100}, [\href{http://arxiv.org/abs/hep-th/9702198}{{\tt
  hep-th/9702198}}].

\bibitem{Jefferson:2017ahm}
P.~Jefferson, H.-C. Kim, C.~Vafa and G.~Zafrir, \emph{{Towards Classification
  of 5d SCFTs: Single Gauge Node}},
  \href{http://arxiv.org/abs/1705.05836}{{\tt 1705.05836}}.

\bibitem{Jefferson:2018irk}
P.~Jefferson, S.~Katz, H.-C. Kim and C.~Vafa, \emph{{On Geometric
  Classification of 5d SCFTs}},
  \href{http://dx.doi.org/10.1007/JHEP04(2018)103}{\emph{JHEP} {\bf 04} (2018)
  103}, [\href{http://arxiv.org/abs/1801.04036}{{\tt 1801.04036}}].

\bibitem{Xie:2017pfl}
D.~Xie and S.-T. Yau, \emph{{Three dimensional canonical singularity and five
  dimensional $ \mathcal{N} $ = 1 SCFT}},
  \href{http://dx.doi.org/10.1007/JHEP06(2017)134}{\emph{JHEP} {\bf 06} (2017)
  134}, [\href{http://arxiv.org/abs/1704.00799}{{\tt 1704.00799}}].

\bibitem{Aharony:1997ju}
O.~Aharony and A.~Hanany, \emph{{Branes, superpotentials and superconformal
  fixed points}},
  \href{http://dx.doi.org/10.1016/S0550-3213(97)00472-0}{\emph{Nucl. Phys.}
  {\bf B504} (1997) 239--271}, [\href{http://arxiv.org/abs/hep-th/9704170}{{\tt
  hep-th/9704170}}].

\bibitem{Kol:1997fv}
B.~Kol, \emph{{5-D field theories and M theory}},
  \href{http://dx.doi.org/10.1088/1126-6708/1999/11/026}{\emph{JHEP} {\bf 11}
  (1999) 026}, [\href{http://arxiv.org/abs/hep-th/9705031}{{\tt
  hep-th/9705031}}].

\bibitem{Aharony:1997bh}
O.~Aharony, A.~Hanany and B.~Kol, \emph{{Webs of (p,q) five-branes,
  five-dimensional field theories and grid diagrams}},
  \href{http://dx.doi.org/10.1088/1126-6708/1998/01/002}{\emph{JHEP} {\bf 01}
  (1998) 002}, [\href{http://arxiv.org/abs/hep-th/9710116}{{\tt
  hep-th/9710116}}].

\bibitem{DeWolfe:1999hj}
O.~DeWolfe, A.~Hanany, A.~Iqbal and E.~Katz, \emph{{Five-branes, seven-branes
  and five-dimensional E(n) field theories}},
  \href{http://dx.doi.org/10.1088/1126-6708/1999/03/006}{\emph{JHEP} {\bf 03}
  (1999) 006}, [\href{http://arxiv.org/abs/hep-th/9902179}{{\tt
  hep-th/9902179}}].

\bibitem{D'Hoker:2016rdq}
E.~D'Hoker, M.~Gutperle, A.~Karch and C.~F. Uhlemann, \emph{{Warped
  $AdS_6\times S^2$ in Type IIB supergravity I: Local solutions}},
  \href{http://dx.doi.org/10.1007/JHEP08(2016)046}{\emph{JHEP} {\bf 08} (2016)
  046}, [\href{http://arxiv.org/abs/1606.01254}{{\tt 1606.01254}}].

\bibitem{DHoker:2016ysh}
E.~D'Hoker, M.~Gutperle and C.~F. Uhlemann, \emph{{Holographic duals for
  five-dimensional superconformal quantum field theories}},
  \href{http://dx.doi.org/10.1103/PhysRevLett.118.101601}{\emph{Phys. Rev.
  Lett.} {\bf 118} (2017) 101601}, [\href{http://arxiv.org/abs/1611.09411}{{\tt
  1611.09411}}].

\bibitem{DHoker:2017mds}
E.~D'Hoker, M.~Gutperle and C.~F. Uhlemann, \emph{{Warped $AdS_6\times S^2$ in
  Type IIB supergravity II: Global solutions and five-brane webs}},
  \href{http://dx.doi.org/10.1007/JHEP05(2017)131}{\emph{JHEP} {\bf 05} (2017)
  131}, [\href{http://arxiv.org/abs/1703.08186}{{\tt 1703.08186}}].

\bibitem{Apruzzi:2014qva}
F.~Apruzzi, M.~Fazzi, A.~Passias, D.~Rosa and A.~Tomasiello, \emph{{AdS$_{6}$
  solutions of type II supergravity}},
  \href{http://dx.doi.org/10.1007/JHEP11(2014)099,
  10.1007/JHEP05(2015)012}{\emph{JHEP} {\bf 11} (2014) 099},
  [\href{http://arxiv.org/abs/1406.0852}{{\tt 1406.0852}}].

\bibitem{Kim:2015hya}
H.~Kim, N.~Kim and M.~Suh, \emph{{Supersymmetric AdS$_6$ Solutions of Type IIB
  Supergravity}},
  \href{http://dx.doi.org/10.1140/epjc/s10052-015-3705-1}{\emph{Eur. Phys. J.}
  {\bf C75} (2015) 484}, [\href{http://arxiv.org/abs/1506.05480}{{\tt
  1506.05480}}].

\bibitem{Kim:2016rhs}
H.~Kim and N.~Kim, \emph{{Comments on the symmetry of AdS$_6$ solutions in
  string/M-theory and Killing spinor equations}},
  \href{http://dx.doi.org/10.1016/j.physletb.2016.07.070}{\emph{Phys. Lett.}
  {\bf B760} (2016) 780--787}, [\href{http://arxiv.org/abs/1604.07987}{{\tt
  1604.07987}}].

\bibitem{Brandhuber:1999np}
A.~Brandhuber and Y.~Oz, \emph{{The D-4 - D-8 brane system and five-dimensional
  fixed points}},
  \href{http://dx.doi.org/10.1016/S0370-2693(99)00763-7}{\emph{Phys. Lett.}
  {\bf B460} (1999) 307--312}, [\href{http://arxiv.org/abs/hep-th/9905148}{{\tt
  hep-th/9905148}}].

\bibitem{Bergman:2012kr}
O.~Bergman and D.~Rodriguez-Gomez, \emph{{5d quivers and their AdS(6) duals}},
  \href{http://dx.doi.org/10.1007/JHEP07(2012)171}{\emph{JHEP} {\bf 07} (2012)
  171}, [\href{http://arxiv.org/abs/1206.3503}{{\tt 1206.3503}}].

\bibitem{Jafferis:2012iv}
D.~L. Jafferis and S.~S. Pufu, \emph{{Exact results for five-dimensional
  superconformal field theories with gravity duals}},
  \href{http://dx.doi.org/10.1007/JHEP05(2014)032}{\emph{JHEP} {\bf 05} (2014)
  032}, [\href{http://arxiv.org/abs/1207.4359}{{\tt 1207.4359}}].

\bibitem{Bergman:2012qh}
O.~Bergman and D.~Rodriguez-Gomez, \emph{{Probing the Higgs branch of 5d fixed
  point theories with dual giant gravitons in AdS(6)}},
  \href{http://dx.doi.org/10.1007/JHEP12(2012)047}{\emph{JHEP} {\bf 12} (2012)
  047}, [\href{http://arxiv.org/abs/1210.0589}{{\tt 1210.0589}}].

\bibitem{Assel:2012nf}
B.~Assel, J.~Estes and M.~Yamazaki, \emph{{Wilson Loops in 5d N=1 SCFTs and
  AdS/CFT}}, \href{http://dx.doi.org/10.1007/s00023-013-0249-5}{\emph{Annales
  Henri Poincare} {\bf 15} (2014) 589--632},
  [\href{http://arxiv.org/abs/1212.1202}{{\tt 1212.1202}}].

\bibitem{Bergman:2013koa}
O.~Bergman, D.~Rodr\'iguez-G\'omez and G.~Zafrir, \emph{{5d superconformal
  indices at large N and holography}},
  \href{http://dx.doi.org/10.1007/JHEP08(2013)081}{\emph{JHEP} {\bf 08} (2013)
  081}, [\href{http://arxiv.org/abs/1305.6870}{{\tt 1305.6870}}].

\bibitem{Pini:2014bea}
A.~Pini and D.~Rodr\'iguez-G\'omez, \emph{{Gauge/gravity duality and RG flows
  in 5d gauge theories}},
  \href{http://dx.doi.org/10.1016/j.nuclphysb.2014.05.009}{\emph{Nucl. Phys.}
  {\bf B884} (2014) 612--631}, [\href{http://arxiv.org/abs/1402.6155}{{\tt
  1402.6155}}].

\bibitem{Chang:2017mxc}
C.-M. Chang, M.~Fluder, Y.-H. Lin and Y.~Wang, \emph{{Romans Supergravity from
  Five-Dimensional Holograms}},
  \href{http://dx.doi.org/10.1007/JHEP05(2018)039}{\emph{JHEP} {\bf 05} (2018)
  039}, [\href{http://arxiv.org/abs/1712.10313}{{\tt 1712.10313}}].

\bibitem{Passias:2018swc}
A.~Passias and P.~Richmond, \emph{{Perturbing AdS$_6 \times_w S^4$: linearised
  equations and spin-2 spectrum}},  \href{http://arxiv.org/abs/1804.09728}{{\tt
  1804.09728}}.

\bibitem{Lozano:2012au}
Y.~Lozano, E.~\'O~Colg\'ain, D.~Rodr\'iguez-G\'omez and K.~Sfetsos,
  \emph{{Supersymmetric $AdS_6$ via T Duality}},
  \href{http://dx.doi.org/10.1103/PhysRevLett.110.231601}{\emph{Phys. Rev.
  Lett.} {\bf 110} (2013) 231601}, [\href{http://arxiv.org/abs/1212.1043}{{\tt
  1212.1043}}].

\bibitem{Lozano:2013oma}
Y.~Lozano, E.~O. O~Colg\'ain and D.~Rodriguez-Gomez, \emph{{Hints of 5d Fixed
  Point Theories from Non-Abelian T-duality}},
  \href{http://dx.doi.org/10.1007/JHEP05(2014)009}{\emph{JHEP} {\bf 05} (2014)
  009}, [\href{http://arxiv.org/abs/1311.4842}{{\tt 1311.4842}}].

\bibitem{Gutperle:2017tjo}
M.~Gutperle, C.~Marasinou, A.~Trivella and C.~F. Uhlemann, \emph{{Entanglement
  entropy vs. free energy in IIB supergravity duals for 5d SCFTs}},
  \href{http://dx.doi.org/10.1007/JHEP09(2017)125}{\emph{JHEP} {\bf 09} (2017)
  125}, [\href{http://arxiv.org/abs/1705.01561}{{\tt 1705.01561}}].

\bibitem{Kaidi:2017bmd}
J.~Kaidi, \emph{{(p,q)-strings probing five-brane webs}},
  \href{http://dx.doi.org/10.1007/JHEP10(2017)087}{\emph{JHEP} {\bf 10} (2017)
  087}, [\href{http://arxiv.org/abs/1708.03404}{{\tt 1708.03404}}].

\bibitem{Gutperle:2018wuk}
M.~Gutperle, C.~F. Uhlemann and O.~Varela, \emph{{Massive spin 2 excitations in
  $AdS_6\times S^2$ warped spacetimes}},
  \href{http://dx.doi.org/10.1007/JHEP07(2018)091}{\emph{JHEP} {\bf 07} (2018)
  091}, [\href{http://arxiv.org/abs/1805.11914}{{\tt 1805.11914}}].

\bibitem{Hong:2018amk}
J.~Hong, J.~T. Liu and D.~R. Mayerson, \emph{{Gauged Six-Dimensional
  Supergravity from Warped IIB Reductions}},
  \href{http://arxiv.org/abs/1808.04301}{{\tt 1808.04301}}.

\bibitem{Malek:2018zcz}
E.~Malek, H.~Samtleben and V.~Vall~Camell, \emph{{Supersymmetric AdS$_{7}$ and
  AdS$_6$ vacua and their minimal consistent truncations from exceptional field
  theory}},  \href{http://arxiv.org/abs/1808.05597}{{\tt 1808.05597}}.

\bibitem{Bergman:2018hin}
O.~Bergman, D.~Rodriguez-Gomez and C.~F. Uhlemann, \emph{{Testing
  AdS$_{6}$/CFT$_{5}$ in Type IIB with stringy operators}},
  \href{http://dx.doi.org/10.1007/JHEP08(2018)127}{\emph{JHEP} {\bf 08} (2018)
  127}, [\href{http://arxiv.org/abs/1806.07898}{{\tt 1806.07898}}].

\bibitem{Fluder:2018chf}
M.~Fluder and C.~F. Uhlemann, \emph{{Precision test of AdS$_6$/CFT$_5$ in Type
  IIB}},  \href{http://arxiv.org/abs/1806.08374}{{\tt 1806.08374}}.

\bibitem{DHoker:2017zwj}
E.~D'Hoker, M.~Gutperle and C.~F. Uhlemann, \emph{{Warped $AdS_6\times S^2$ in
  Type IIB supergravity III: Global solutions with seven-branes}},
  \href{http://dx.doi.org/10.1007/JHEP11(2017)200}{\emph{JHEP} {\bf 11} (2017)
  200}, [\href{http://arxiv.org/abs/1706.00433}{{\tt 1706.00433}}].

\bibitem{Gutperle:2018vdd}
M.~Gutperle, A.~Trivella and C.~F. Uhlemann, \emph{{Type IIB 7-branes in warped
  $AdS_6$: partition functions, brane webs and probe limit}},
  \href{http://arxiv.org/abs/1802.07274}{{\tt 1802.07274}}.

\bibitem{Benini:2009gi}
F.~Benini, S.~Benvenuti and Y.~Tachikawa, \emph{{Webs of five-branes and N=2
  superconformal field theories}},
  \href{http://dx.doi.org/10.1088/1126-6708/2009/09/052}{\emph{JHEP} {\bf 09}
  (2009) 052}, [\href{http://arxiv.org/abs/0906.0359}{{\tt 0906.0359}}].

\bibitem{Lozano:2018pcp}
Y.~Lozano, N.~T. Macpherson and J.~Montero, \emph{{$AdS_6$ T-duals and Type IIB
  $AdS_6\times S^2$ Geometries with 7-Branes}},
  \href{http://arxiv.org/abs/1810.08093}{{\tt 1810.08093}}.

\bibitem{Chacaltana:2010ks}
O.~Chacaltana and J.~Distler, \emph{{Tinkertoys for Gaiotto Duality}},
  \href{http://dx.doi.org/10.1007/JHEP11(2010)099}{\emph{JHEP} {\bf 11} (2010)
  099}, [\href{http://arxiv.org/abs/1008.5203}{{\tt 1008.5203}}].

\bibitem{Bergman:2014kza}
O.~Bergman and G.~Zafrir, \emph{{Lifting 4d dualities to 5d}},
  \href{http://dx.doi.org/10.1007/JHEP04(2015)141}{\emph{JHEP} {\bf 04} (2015)
  141}, [\href{http://arxiv.org/abs/1410.2806}{{\tt 1410.2806}}].

\bibitem{Hayashi:2014hfa}
H.~Hayashi, Y.~Tachikawa and K.~Yonekura, \emph{{Mass-deformed T$_{N}$ as a
  linear quiver}}, \href{http://dx.doi.org/10.1007/JHEP02(2015)089}{\emph{JHEP}
  {\bf 02} (2015) 089}, [\href{http://arxiv.org/abs/1410.6868}{{\tt
  1410.6868}}].

\bibitem{Tachikawa:2015bga}
Y.~Tachikawa, \emph{{A review of the $T_N$ theory and its cousins}},
  \href{http://dx.doi.org/10.1093/ptep/ptv098}{\emph{PTEP} {\bf 2015} (2015)
  11B102}, [\href{http://arxiv.org/abs/1504.01481}{{\tt 1504.01481}}].

\end{thebibliography}\endgroup
\end{document}